\documentclass{pazhastl}

\usepackage{apjfonts}

\usepackage{natbib}
\usepackage{graphicx}
\usepackage[hyperfootnotes=false]{hyperref}

\def\smfigure#1#2#3{
  \begin{minipage}{1.0\columnwidth}
    \begin{minipage}{0.049\columnwidth}
      \rotatebox{90}{\small\phantom{0000}#3}
    \end{minipage}
    \begin{minipage}{0.9\columnwidth}
      \includegraphics[bb=40 188 556 678,width=0.97\columnwidth]{#1}
      \centerline{\small #2}
    \end{minipage}

    \vskip 3pt
    ~
  \end{minipage}
}

\def\smfiguresmall#1#2#3{
  \begin{minipage}{0.95\columnwidth}
    \begin{minipage}{0.049\columnwidth}
      \rotatebox{90}{\footnotesize\phantom{0000}#3}
    \end{minipage}
    \begin{minipage}{0.95\columnwidth}
      \includegraphics[bb=20 188 576 540,width=0.97\columnwidth]{#1}
      \centerline{\footnotesize #2}
    \end{minipage}

    \vskip 3pt
    ~
  \end{minipage}
}

\begin{document}

\journalinfo{2011}{37}{2}{0}[0]

\title{Fast optical variability of SS\,433}

\author{R.~A.~Burenin\email{rodion@hea.iki.rssi.ru}\address{1},
M.~G.~Revnivtsev\address{1,2},
I.~M.~Khamitov\address{3},
I.~F.~Bikmaev\address{4,5},
A.~S.~Nosov\address{1},
M.~N.~Pavlinsky\address{1},
R.~A.~Sunyaev\address{1,6}
\addresstext{1}{Space Research Institute (IKI), Moscow, Russia}
\addresstext{2}{Excellence Cluster Universe, Technische Universit\"at
  M\"unchen, Germany }
\addresstext{3}{TUBITAK National Observatory, Antalya, Turkey}
\addresstext{4}{Kazan (Volga Region) State University, Kazan, Russia}
\addresstext{5}{Academy of Sciences of Tatarstan, Kazan, Russia}
\addresstext{6}{Max-Planck-Institut f\"ur Astrophysik, Garching, Germany}
}

\shortauthor{Burenin et al.}

\shorttitle{Fast optical variability of SS\,433}

\submitted{April 13, 2010}

\begin{abstract}
  We study the optical variability of the peculiar Galactic source
  SS\,433 using the observations made with the Russian Turkish 1.5-m
  telescope (RTT150). A simple technique which allows to obtain
  high-quality photometric measurements with 0.3--1~s time resolution
  using ordinary CCD is described in detail. Using the test
  observations of nonvariable stars, we show that the atmospheric
  turbulence introduces no significant distortions into the measured
  light curves. Therefore, the data obtained in this way are well
  suited for studying the \emph{aperiodic} variability of various
  objects.

  The large amount of SS\,433 optical light curve measurements
  obtained in this way allowed us to obtain the power spectra of its
  flux variability with a record sensitivity up to frequencies of
  $\sim0.5$~Hz and to detect its break at frequency $\approx
  2.4\times10^{-3}$~Hz. We suggest that this break in the power
  spectrum results from the smoothing of the optical flux variability
  due to a finite size of the emitting region. Based on our
  measurement of the break frequency in the power spectrum, we
  estimated the size of the accretion-disk photosphere as $2\times
  10^{12}$~cm. We show that the amplitude of the variability in
  SS\,433 decreases sharply during accretion-disk eclipses, but it
  does not disappear completely. This suggests that the size of the
  variable optical emission source is comparable to that of the normal
  star whose size is therefore $R_O\approx 2\times 10^{12}$~cm
  $\approx30 R_\odot$. The decrease in flux variability amplitude
  during eclipses suggests the presence of a nonvariable optical emission
  component with a magnitude $m_R\approx13.2$.



\end{abstract}

\section{Introduction}
\label{sec:intro}

The object SS\,433 is a binary system with coninuous accretion onto a
compact object, most probably a black hole \citep[for a review see,
e.g.,][]{fabrika04}. It is the only known object of this kind in
Galaxy and similar objects seen face-on are probably observed as
ultraluminous X-ray sources in other galaxies
\citep{fm01,king02,begelman06}.

The emission from SS\,433 is variable at all observed time scales and
at all wavelengths of the electromagnetic spectrum \citep[see,
e.g.,][]{cherepaschuk81,ggch83,1987MNRAS.228..293S,gech98,2001ApJ...561.1027E,cherepaschuk05,kotani06,trushkin07}.
Several types of periodic variability were detected in the variability
of the source: precessional, orbital, and nutational; this allowed the
parameters of the binary system to be constrained significantly
\citep{cherepaschuk02}. The studies of eclipses also allowed to
estimate the geometrical sizes of the various components of the binary
system \citep[see, e.g.,][]{1987MNRAS.228..293S}.

The aperiodic variability of SS\,433 was also studied
\citep{mikej04,mikej06}. It was shown that the the broadband power
spectrum of SS\,433 is a power law with a break at a frequency of
about $10^{-7}$~Hz below which the power spectrum is flat. Such a
power spectrum is expected in the model of self-similar production of
accretion-rate variability in an accretion disk, and the spectrum
flatness at low frequencies is explained by the fact that no
additional variability can be produced on time scales longer than the
viscous time scale in accretion disk
\citep{lyubarsky97,churazov01,ga05}.

Currently, the variability of SS\,433 is poorly studied at higher
frequencies $\ga10^{-3}$--$10^{-2}$~Hz \citep{fabrika04,mikej06}. On
the other hand, these frequencies should correspond to the size of
X-ray and optical emission regions in the object and it should take
its effect in the variability of the source at these high frequences.
In this paper we present the results of our study of optical
variability of SS\,433 at high frequencies, using a significant amount
of new data on fast optical photometry of SS\,433 with $\approx 1$~s
time resolution.

\begin{figure*}
  \centering
  \includegraphics[width=0.65\linewidth]{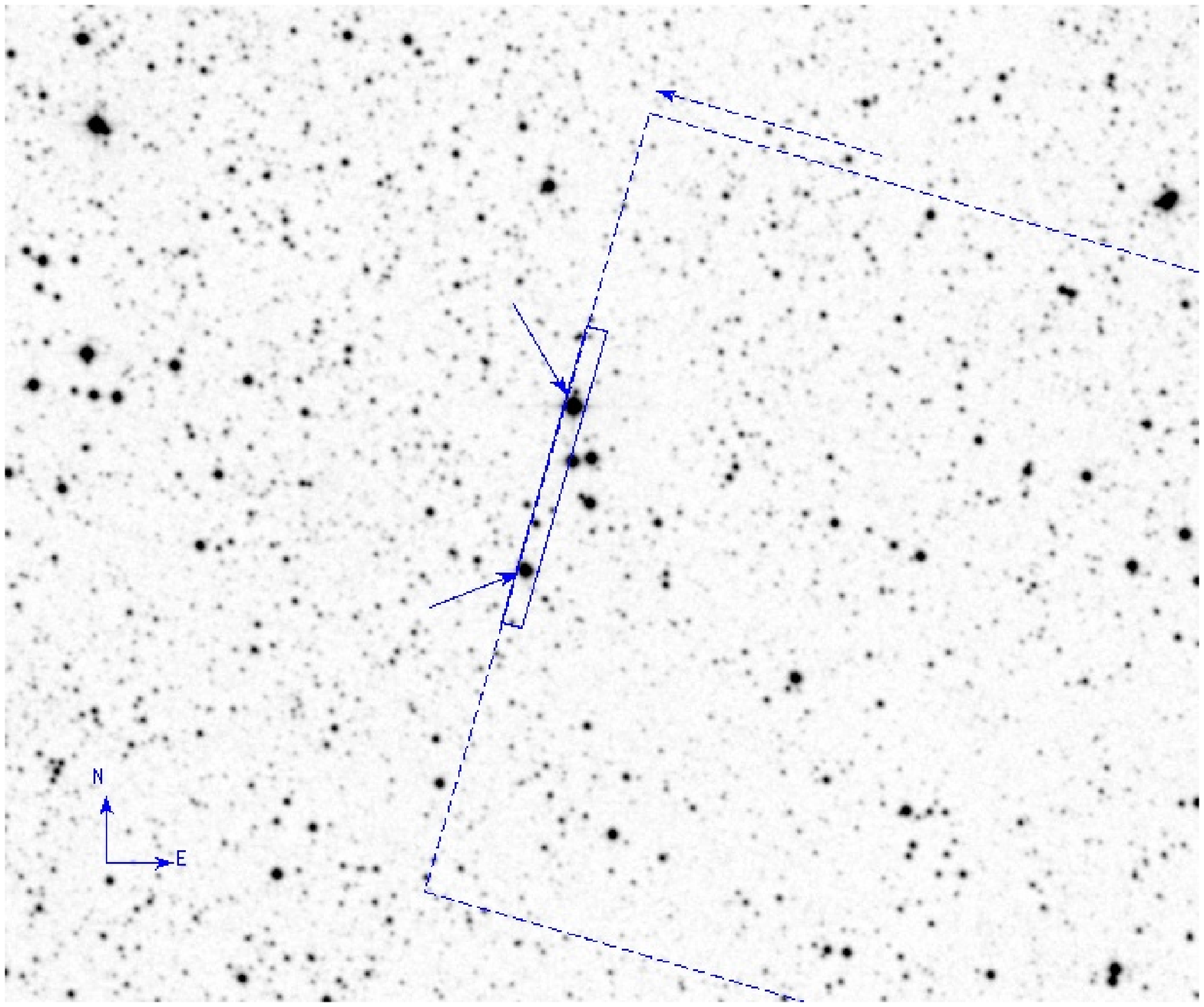}

  \bigskip

  \includegraphics[width=0.65\linewidth]{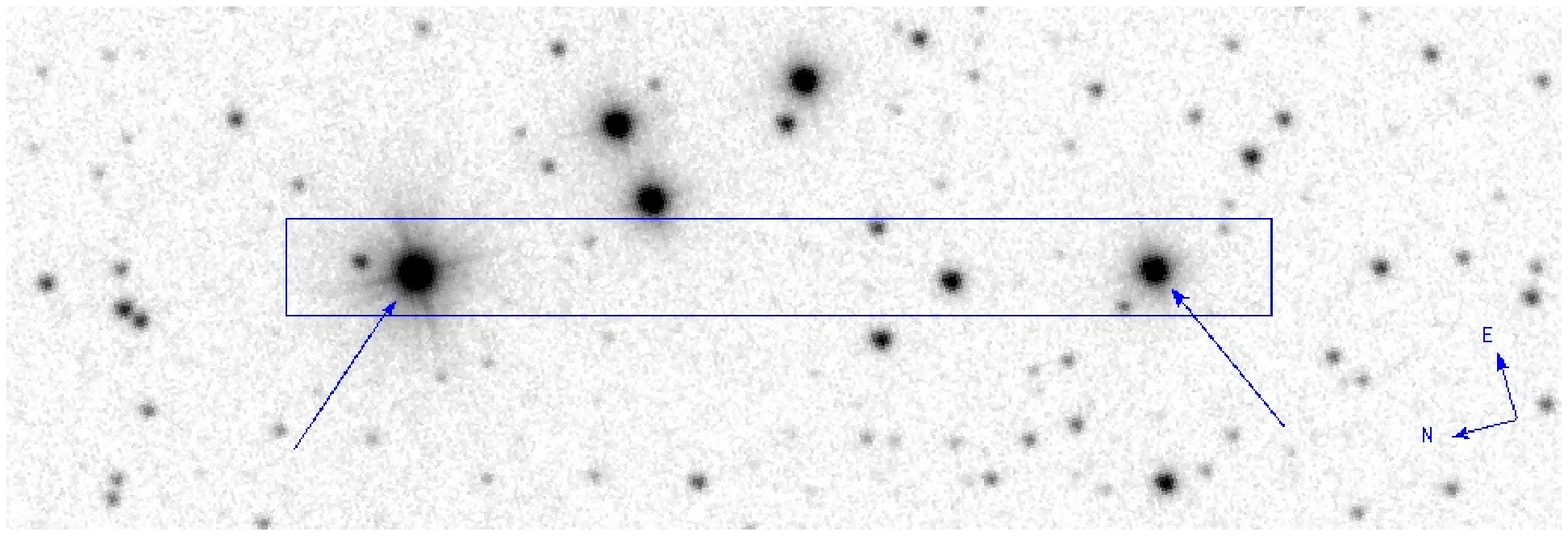}

  \hskip -3mm
  \includegraphics[bb=25 187 573 376,width=0.65\linewidth]{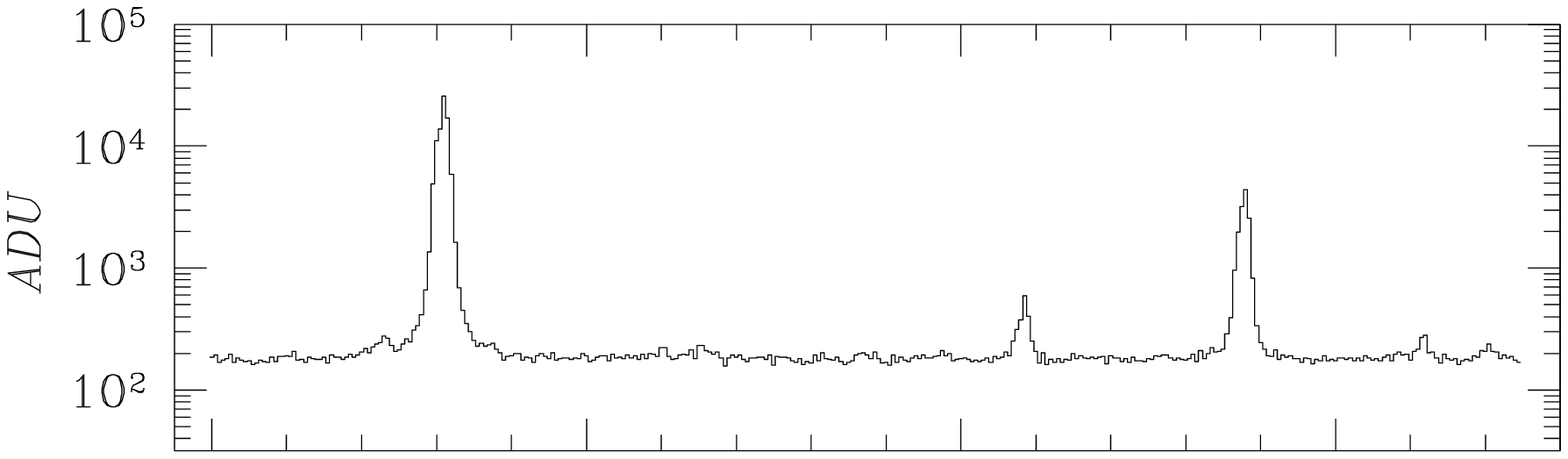}
  
    
  
  \caption{The field near SS\,433. We show here, how the telescope was
    pointed to object and what part of the CCD was used during our
    observations. The direction of CCD readout is shown with dashed
    arrow in the upper panel. The object and the reference star are
    also shown with arrows. In the lower panel the example of
    one-dimentional line of the data which was read out after the
    every exposure is shown.}

  \label{fig:field}
\end{figure*}

\section{Observations}
\label{sec:obs}

The observations were carried out with Russian-Turkish 1.5-m telescope
(RTT150)\footnote{http://hea.iki.rssi.ru/rtt150/}, using
CCD-photometer at the Cassegrain focus of the telescope
$f=1/7.7$. Photometrical measurements were done with Andor DW-436 CCD
camera. This is a $2048\times2048$ back-illuminated CCD, cooled
electronically to $-60$~K. This CCD has high quantum efficiency (more
then 90\% in $R$ band), negligibly small dark current and low readout
noise ($\approx 2~e$). At the telescope focus the angular size of CCD
pixel is $0.24$\arcsec, the size of field of view --- about
$8$\arcmin.

\subsection{Instrument Setup}
\label{sec:setup}

In order to reduce the CCD readout time the observations were made as
shown in Fig.~\ref{fig:field}. Since CCD readout speed depends mainly
on linear size across the readout axis, telescope pointing was done so
that the object and the reference star are set in parallel to that
edge of the CCD, where the readout take place. The shutter was open
during all the observations; therefore, both stars were set close to
the readout side to avoid data contamination by bright stars outside
the cutout strip.

\begin{figure*}
  \centering
  \smfiguresmall{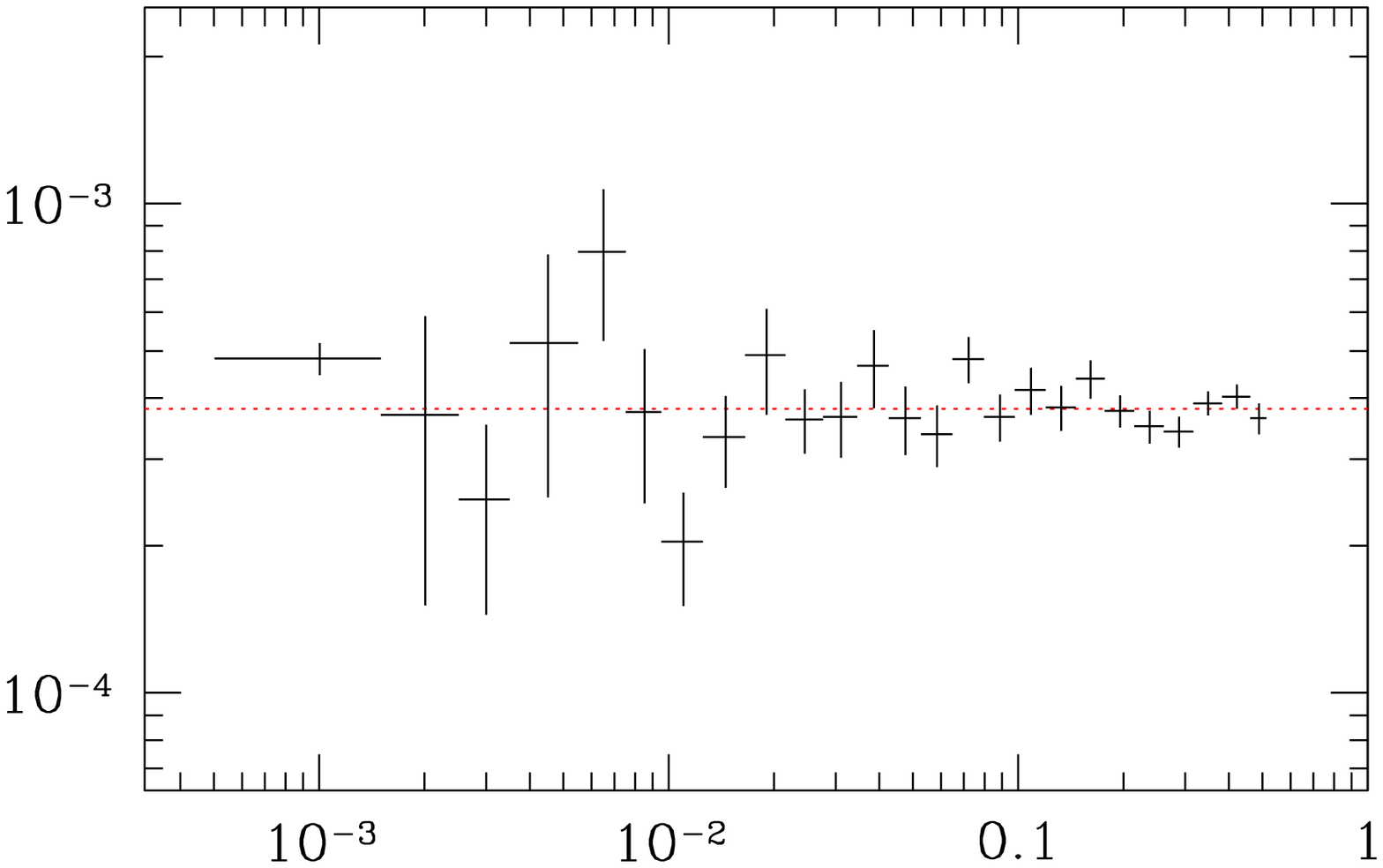}{Frequency, Hz}{Power,
    $(\sigma_x/\langle x \rangle)^2 /$~Hz}
  ~~~
  \smfiguresmall{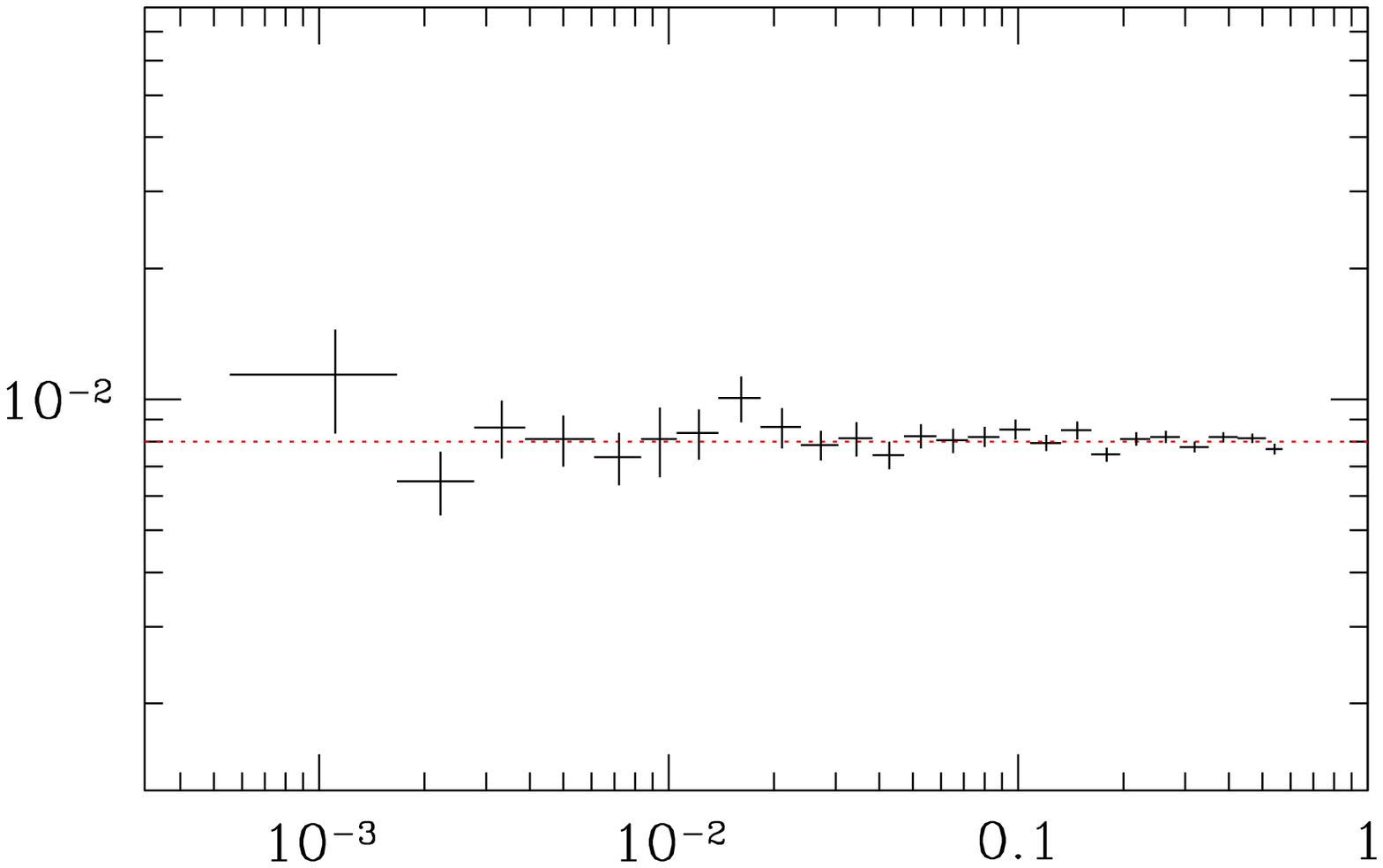}{Frequency, Hz}{Power,
    $(\sigma_x/\langle x \rangle)^2 /$~Hz}

  \caption{Power spectra of nonvariable stars. In the left panel ---
    the results of the observations of nonvariable stars in the same
    nights when SS\,433 was observed, in the right panel --- the
    results of observations carried out later with slightly different
    signal-to-noise ratio.}
  \label{fig:powspec_stars}
\end{figure*}

Only the narrow CCD strip with the object and the reference star was
read out during the observations. The strip width was chosen so that
the main parts of the wings of the point spread function (PSF) were
observed inside this strip. This strip did not change from observation
to observation and its width was set to be 12\arcsec (50 CCD
pixels). The telescope was pointed so that these two stars were set at
the center of the cutout CCD strip as precisely as possible; then the
autoguider was immediately switched on. Since the telescope guiding
accuracy is a few tenths of an arcsecond, one can be sure that both
stars during the observations remained at the center of the cutout
strip with good accuracy.

The CCD strip was read out not as an image but as a one-dimensional
data line by binning all 50 CCD pixels of the strip width into single
data pixel (Fig.~\ref{fig:field}). In addition, the CCD was also
binned by two pixels along the length of the cutout strip. This
allowed to reduce significantly the readout time and readout noise,
while the sky background was still much lower than the flux from the
object. In this setup the readout time of our CCD was about 0.3~s.

The exposure time was set to be about 0.7~s so that the exposure and
readout cycle was exactly 0.974 s during all the observations. The
data were initially written to computer random access memory and only
every 1000 measurements were then written to the hard disk. This
allowed to avoid the possible delays associated with the data writing
to the hard disk within blocks of 1000 measurements.

The absolute timing marker was also placed in the data at the instant
of their writing to the hard disk. Due to technical difficulties
during our observations no accurate data timing based on the GPS
signal was used. Instead, the computer clock was synchronized with the
GPS signal every evening before and during the observations; the time
markers obtained from this clock were then used. Thus, the absolute
data timing accuracy should not be worse than a few tenths of a
second.

It is important that our observing technique allow to obtain
simultaneous measurements of the fluxes of at least two stars, the
reference star and the object. This makes it possible to perform
differential photometric measurements, which allows to take into
account the extinction variations at different zenith distances and
due to light clouds. In addition, as we show below, this also allows
to take into account the most of the influence of the variable
absorption and stellar jitter due to atmospheric turbulence up to
frequencies of $\sim0.5$~Hz.

The fluxes were measured inside the linear aperture centered on the
profile of the signal from the star (Fig.~\ref{fig:field}). The window
size was set for each segment of 1000 measurements separately at six
RMS widths of the PSF. Thus, no more than a few tenths of a percent of
the stellar flux was observed outside the aperture. The background was
fitted by linear function in each strip being read out separately;
during this procedure the stellar flux was eliminated by applying the
standard sigma-clipping algorithm.

\subsection{The Influence of Atmospheric Turbulence on Photometric
  Measurements}
\label{sec:atm}

In order to study the \emph{aperiodic} variability of various objects,
the properties of the noise that emerges during the measurements of
object fluxes must be well known. The true shape of spectrum of the
object's variable emission will not be distorted if the flux
measurement errors are independent in each measurement. In this case,
the noise power spectrum is a constant that can be subtracted from the
power spectrum of the measured emission of the object in order to
obtain the power spectrum of its intrinsic variability.

The photon counting noise is definitely independent for different
measurements, but in our case the errors could also emerge for other
reasons. These primarily include the influence of atmospheric
turbulence. There are always chaotic variations in the temperature and
refractive index of the medium due to the presence of turbulence in
the atmosphere. This leads to a distortion of the plane wave front,
which manifests itself as jitter, a change of the shape, and
scintillations of stars. All this, in turn, can affect the results of
photometric measurements, with the corresponding measurement errors
having a complex power spectrum dependent on the pattern of
atmospheric turbulence.

According to the preliminary data of stellar jitter studies during
RTT-150 observations, turbulence emerging on the telescope dome
introduces a significant fraction of the distortions. This turbulence
affects all stars in the telescope field of view in the very same way
and, hence, it should not have any effect on differential photometric
measurements.

Turbulence at high altitudes can also have a significant effect on the
observations. Starting from the altitude of 2.5~km, at which the
TUBITAK Observatory is located, the bulk of the variability of the
refractive index is gained up to an altitude of about 10~km
\citep[see, e.g.,][]{zuev88}. At this altitude, the angular separation
of 1.7\arcmin\ between SS\,433 and the reference star corresponds to a
linear separation of about 3--4~m, which is larger than the telescope
mirror size. Therefore, the light from these stars passes through
different parts of the atmosphere and their jitter can
differ. However, at a wind speed of $\sim10$~m/s, the turbulence will
be essentially averaged out at a time scale of $\sim1$~s in a fixed
region with a size of the order of the telescope mirror
size. Therefore, one may expect that this turbulence will not
contribute strongly to the photometric measurement errors as well.

In order to test this assumption directly, we carried out the
observations of nonvariable stars with the same instrument setup that
was used for the observations of SS\,433. The subsequent data
reduction was also performed in exactly the same way. The derived
power spectra are shown in Fig.~\ref{fig:powspec_stars}. The left
panel of Figure~\ref{fig:powspec_stars} shows the power spectrum of
the nonvariable stars whose observations were carried out on the same
nights as those of SS\,433. The stars were chosen so that their fluxes
were close to those from SS\,433 and the reference star near this
object. We see that, within the measurement errors, the shape of the
power spectrum is consistent with the assumption that the power does
not depend on frequency. In this case, the value of the constant power
is close to what is obtained during the observations of SS\,433 (see
below).

The right panel of Figure~\ref{fig:powspec_stars} presents the results
of our observations carried out on different nights and with slightly
different CCD settings and stellar flux ratio. Within the errors, the
shape of this power spectrum also agrees with a constant. During these
observations, the signal-to-noise ratio differ from that was obtained
during the observations of SS\,433. This is reflected in a different
value of a constant in the observed power specrtum.  However, all the
uncertainties in the stellar fluxes measurements that can arise from
atmospheric turbulence must be multiplicative (the variations in
atmospheric transparency, the fraction of the flux in the PSF wings,
and so on). Therefore, to estimate the possible distortions of the
power spectrum, only the relative deviations from the constant value
should be considered.

\begin{figure}
  \centering
  \smfigure{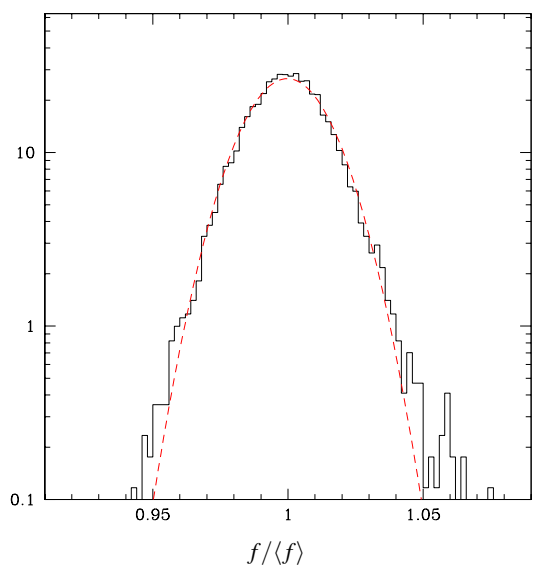}{$f/\langle f\rangle$}{}
  \caption{The distribution of the errors in nonvariable stars flux
    measurements.}
  \label{fig:stars_ds_flux}
\end{figure}

The distribution of errors in the flux measurements of nonvariable
stars is shown in Fig.~\ref{fig:stars_ds_flux}. In these observations,
the stars were chosen so that the signal-to-noise ratio during these
observations was close to that obtained in the observations of
SS\,433. One can see that, on the whole, this distribution is
consistent with a Gaussian one; there are slight differences from the
Gaussian distribution only at deviations $>2\sigma$.

Thus, the flux measurement errors in our observations of SS\,433 are
independent and Gaussian with a good accuracy. Therefore, the power
spectrum of the object's intrinsic variability can be properly
measured by subtracting the constant power which depends on the flux
measurement errors during the observations.

\begin{table*}
  \centering
  \caption{Observation Log}
  \label{tab:obs}
  \medskip
  \begin{tabular}{lcccccc}
    \hline
    \hline
    Date &  Start & End & $\Delta t$, & \multicolumn{3}{c}{Phase, $\varphi$} \\
    yymmdd & \multicolumn{2}{c}{\emph{MJD}} & s & prec. & orb. & nut.  \\
    \hline
    040830 & 53247.9597 & 53247.9929 & 0.393 & 0.99 & 0.51 & 0.83 \\
    040831 & 53248.8927 & 53248.9947 & 0.393 & 1.00 & 0.58 & 0.99 \\
    040901 & 53249.8724 & 53249.9653 & 0.393 & 1.00 & 0.66 & 0.14 \\
    050612 & 53533.9452 & 53533.9525 & 1.273 & 0.75 & 0.37 & 0.31 \\
    050614 & 53535.9376 & 53536.0348 & 1.009 & 0.76 & 0.52 & 0.64 \\
    050615 & 53536.9746 & 53537.0547 & 1.275 & 0.77 & 0.60 & 0.80 \\
    050616 & 53537.9440 & 53538.0480 & 0.975 & 0.78 & 0.68 & 0.96 \\
    050617 & 53538.9852 & 53539.0754 & 0.974 & 0.78 & 0.76 & 0.12 \\
    050618 & 53539.9572 & 53540.0797 & 0.974 & 0.79 & 0.83 & 0.28 \\
    050619 & 53540.9113 & 53540.9917 & 0.974 & 0.80 & 0.90 & 0.43 \\
    050620 & 53541.9255 & 53542.0343 & 0.974 & 0.80 & 0.98 & 0.59 \\
    050621 & 53542.9881 & 53543.0764 & 0.974 & 0.81 & 0.06 & 0.76 \\
    050622 & 53543.8850 & 53544.0576 & 0.974 & 0.81 & 0.13 & 0.91 \\
    050719 & 53570.8337 & 53571.0766 & 0.974 & 0.98 & 0.20 & 0.20 \\
    050723 & 53574.8387 & 53575.0375 & 0.974 & 1.00 & 0.50 & 0.83 \\
    050724 & 53575.8974 & 53576.0754 & 0.974 & 0.01 & 0.58 & 1.00 \\
    050725 & 53576.9456 & 53577.0697 & 0.974 & 0.02 & 0.66 & 0.16 \\
    050726 & 53577.8752 & 53578.0444 & 0.974 & 0.02 & 0.73 & 0.31 \\
    050727 & 53578.9636 & 53579.0651 & 0.974 & 0.03 & 0.81 & 0.48 \\
    050728 & 53579.9470 & 53580.0485 & 0.974 & 0.04 & 0.89 & 0.64 \\
    050729 & 53580.8228 & 53580.9920 & 0.974 & 0.04 & 0.96 & 0.78 \\
    050730 & 53581.8715 & 53581.9959 & 0.974 & 0.05 & 0.04 & 0.95 \\
    050731 & 53582.8602 & 53582.9989 & 0.974 & 0.05 & 0.11 & 0.10 \\
    050802 & 53584.9164 & 53584.9763 & 0.974 & 0.07 & 0.27 & 0.42 \\
    050804 & 53586.8539 & 53586.9369 & 0.974 & 0.08 & 0.42 & 0.73 \\
    051014 & 53657.8000 & 53657.8011 & 0.982 & 0.51 & 0.84 & 0.01 \\
    051028 & 53671.7010 & 53671.7236 & 0.974 & 0.60 & 0.90 & 0.22 \\
    \hline
  \end{tabular}
\end{table*}

\begin{figure}
  \centering
  \smfigure{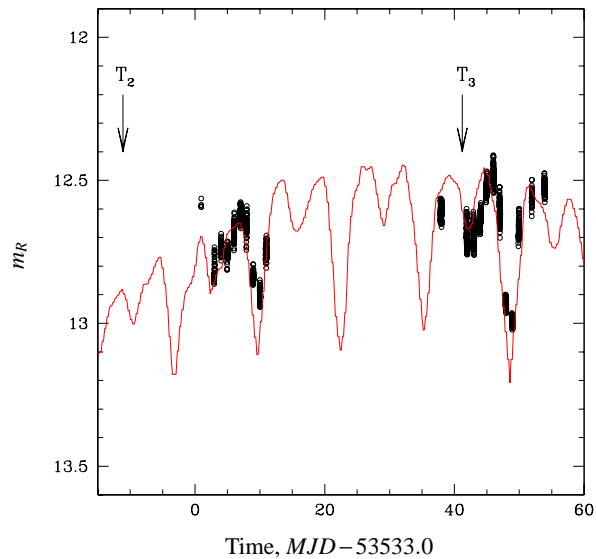}{Time, $MJD-53533.0$}{$m_R$}
  \caption{The points shows the $R$-band magnitudes of SS\,433. Solid
    line shows the mean light curve in $V$ band from the observations
    of \cite{gech98}, taken from the review by \cite{fabrika04}.}
  \label{fig:obs}
\end{figure}

\subsection{The Observations of SS\,433}
\label{sec:obs_ss}

The observations of SS\,433 were carried out at RTT-150 telescope with
CCD photometer, as was described above, mainly in the summer of
2005. In addition, a few observations were also carried out in the
summer of 2004 and the fall of 2005. The observations were performed
in the R band. The detailed data on these observations are presented
in Table~\ref{tab:obs}. The table provides the information on the date
of observations (columns 1, 2, and 3), the time resolution during the
observations ($\Delta t$, column 4), and the precession, orbital, and
nutation phases of SS\,433 at the time of observations (columns 5, 6,
and 7, respectively), according to the ephemerides taken from the
review by \cite{fabrika04}, see also \cite{gech98}.

In Fig.~\ref{fig:obs} the $R$-band magnitude measurements for SS\,433
obtained during our observations in the summer of 2005 are
presented. The solid curve indicates the average $V$-band light curve
constructed from the observations by \cite{gech98} taken from the
review by \cite{fabrika04}. To make this curve coincident with our
data, we set the phase based on the ephemerides of the orbital period
taken from the same review and shifted the magnitudes so that they
roughly coincided with our $R$-band observations. The arrows indicates
the time of maximum of the precessional variability, when the
accretion disk is maximally turned face-on to the observer's direction
($T_3$, $\varphi_{prec}=1$), again according to the ephemerides from
\cite{fabrika04}.

We see from Table~\ref{tab:obs} and Fig.~\ref{fig:obs} that the
observations were carried out mainly close in time to the precession
phase $T_3$. This was done to increase the sensitivity with respect to
the variable emission of the object, because the variability amplitude
is assumed to be highest at this precession phase. The observations
were also performed during disk eclipses --- a total of three eclipses
were observed in the summer and the fall of 2005.

\begin{table}
  \centering
  \caption{The results of our measurements}
  \label{tab:data}
  \medskip
  \begin{tabular}{ccc}
    \hline
    \hline
    \emph{MJD} & $f$ & $\delta f$ \\
    \hline
    53247.959665  &  0.1324 & 0.0220 \\
    53247.959670  &  0.1383 & 0.0229 \\
    53247.959674  &  0.1242 & 0.0218 \\
    53247.959679  &  0.1467 & 0.0214 \\
    53247.959683  &  0.1296 & 0.0215 \\
    53247.959688  &  0.1459 & 0.0218 \\
    53247.959692  &  0.1220 & 0.0213 \\
    53247.959697  &  0.1630 & 0.0239 \\
    53247.959702  &  0.1588 & 0.0228 \\
    53247.959706  &  0.1463 & 0.0230 \\
    \dots & \dots & \dots \\
       \hline
  \end{tabular}

  \medskip

  \begin{minipage}{0.85\linewidth}
    \footnotesize

    \emph{Note:} --- Here only a small part of the table is shown as
    an example. The whole table contains 238638 lines and is available
    in its entirety in the electronic version of the journal, and also
    at: \mbox{http://hea.iki.rssi.ru/rtt150/en/ss433\_pazh10/}

  \end{minipage}
\end{table}

\begin{figure*}
  \centering
  \begin{minipage}{0.8\linewidth}
    \begin{minipage}{0.049\linewidth}
      \rotatebox{90}{\small\phantom{0000} Flux}
    \end{minipage}
    \begin{minipage}{0.9\linewidth}
      \includegraphics[bb=40 188 556 440,width=0.97\linewidth]
      {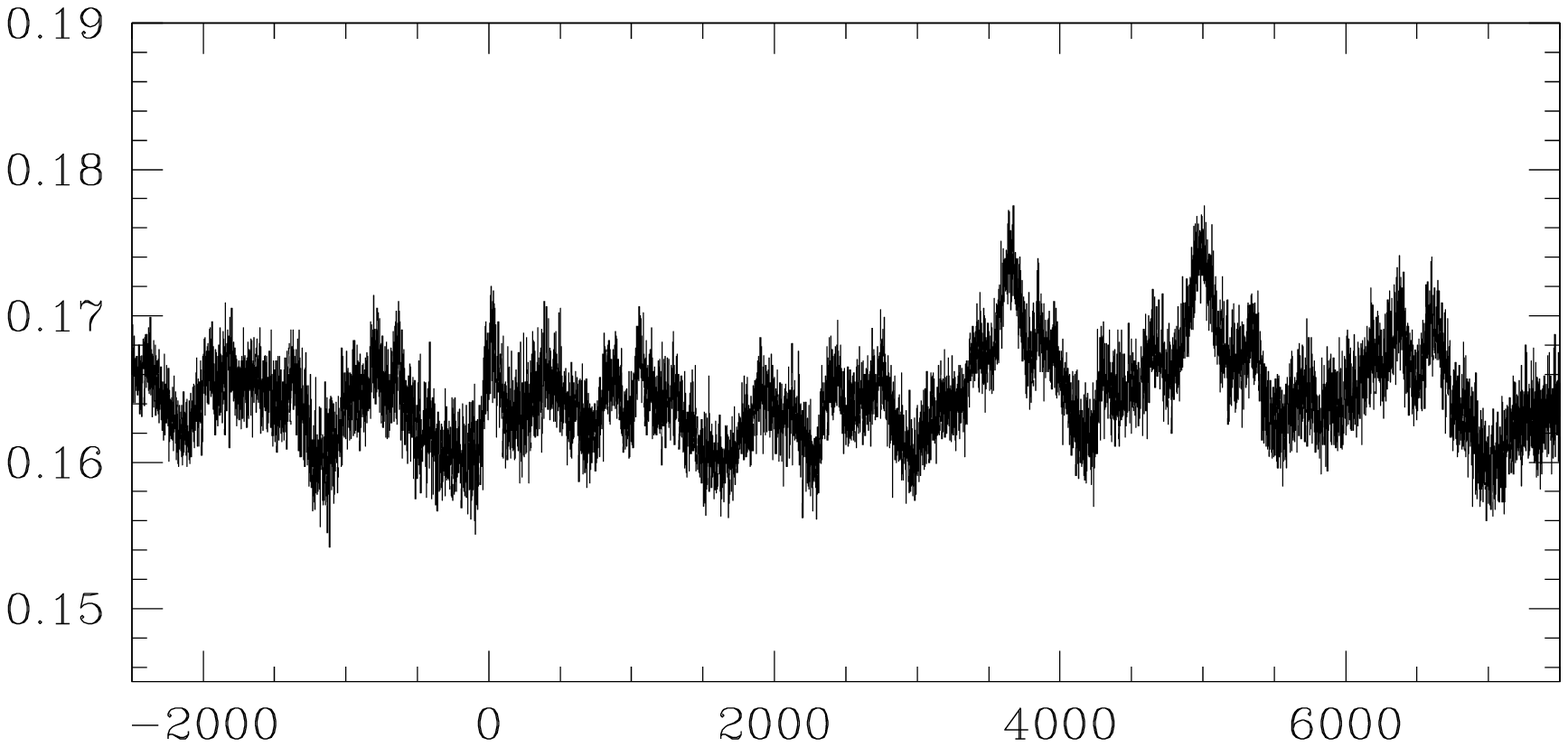}
      \includegraphics[bb=40 188 556 440,width=0.97\linewidth]
      {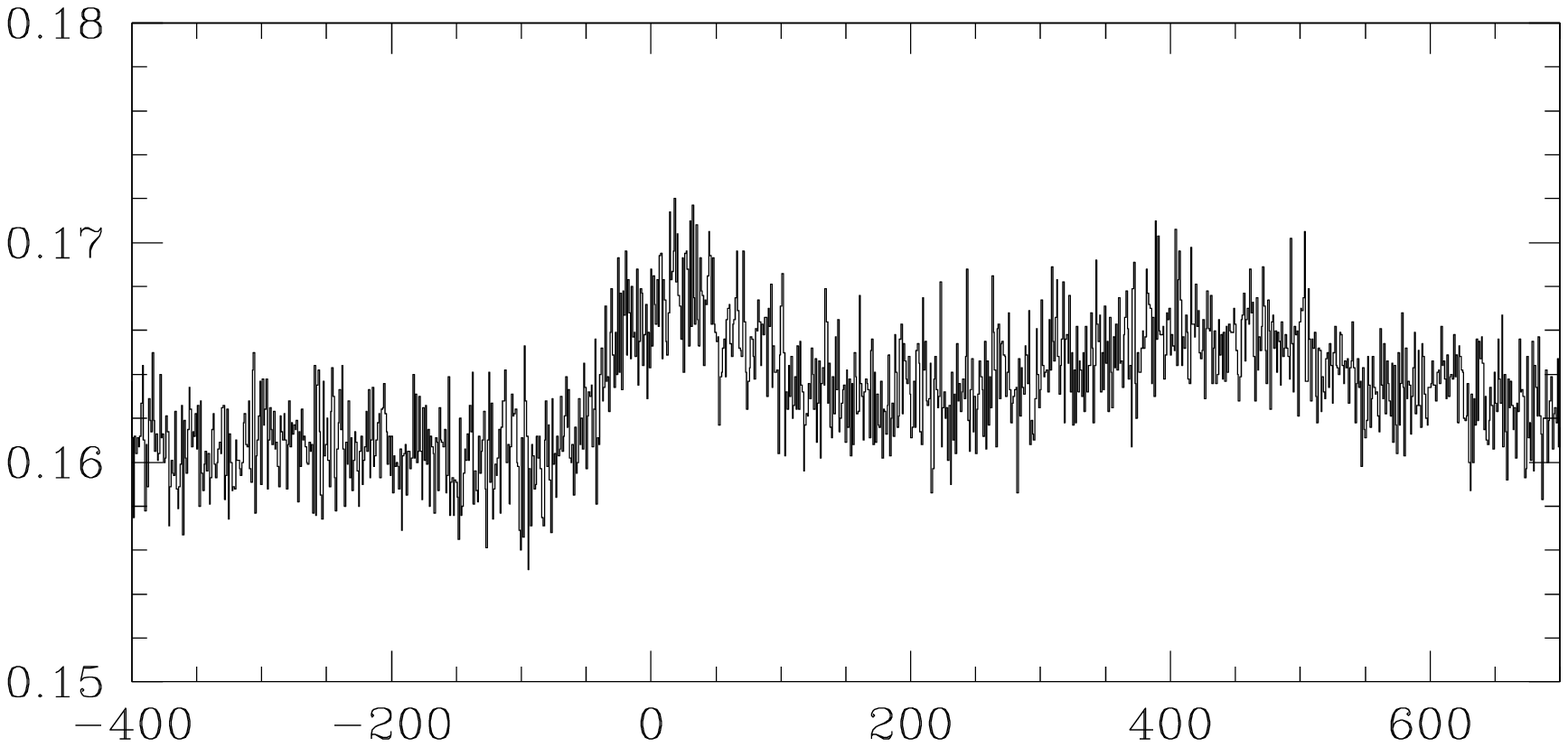}
      \includegraphics[bb=40 188 556 440,width=0.97\linewidth]
      {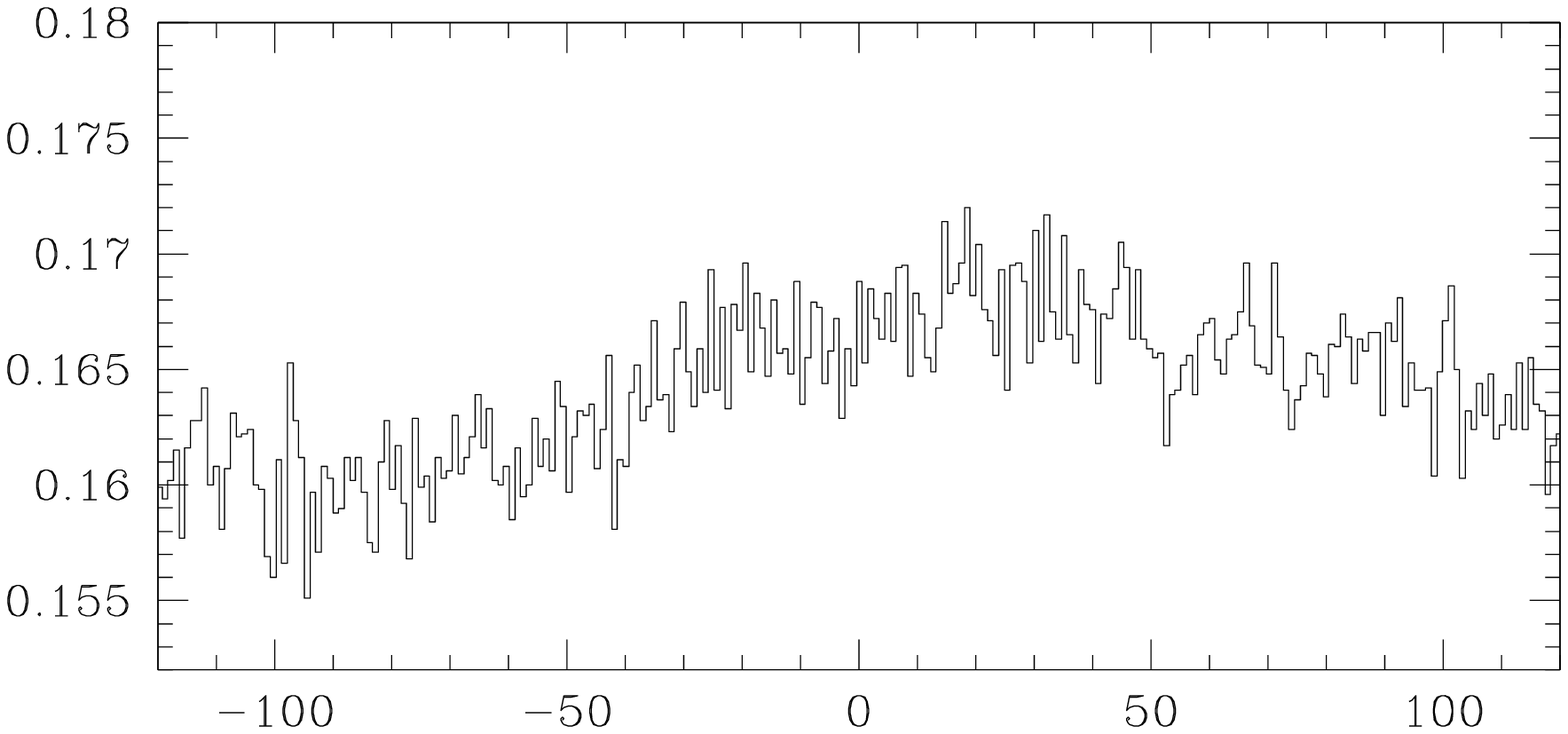}
      \centerline{\small Time, s}
    \end{minipage}

    \vskip 3pt
    ~
  \end{minipage}

  \caption{A segment of the light curve of SS\,433, shown at different
    time scales. The zero time on all panels corresponds to
    \emph{MJD}\,$=53577.92245$.}
  \label{fig:lcex}
\end{figure*}

Due to weather restrictions and because of technical problems, the
observations were not always carried out continuously during the
night. Nevertheless, the bulk of the data are continuous time series
of observations, each with a duration of several hours. In our work,
we used a total of about 240 thousand optical flux measurements for
SS\,433; about 190 thousand are the measurements with a time
resolution of about 1~s obtained during 2005. The results of our
R-band flux measurements for SS\,433 are summarized in
Table~\ref{tab:data}. In Fig.~\ref{fig:lcex} the examples of the
measured SS\,433 light curves with various time resolutions are
shown. Here and everywhere below, unless stated otherwise, the optical
flux from SS\,433 is given in the R band in fractions of the flux of
the reference star (see Fig.~\ref{fig:field}), whose magnitude,
according to our measurements, is $m_R = 10.57$.

The flux measurement errors differ from observation to observation,
because they depend on the sky background, the PSF width determined by
the degree of atmospheric turbulence, and the like. However, their
value is about 2\% of the measured flux in most observations. The
observations performed in the late summer of 2004 are an exception,
since the exposure times in these observations were reduced too much
in order to achieve a higher time resolution.

The errors presented in Table~\ref{tab:data} were determined from the
Poissonian noise of the electrons recorded by the CCD. In order to
take into account the influence of atmospheric jitter and other
effects, this noise was multiplied by the correction factor calculated
from the observations of nonvariable stars, which turned out to be
equal to $\approx2$. Thus, the errors in Table~\ref{tab:data} must be
close to the true measurement errors. They can be used to estimate the
data quality. However, these errors are not used below to study the
variability of the object.

\section{Power spectra}

\begin{figure}
  \centering
  \smfigure{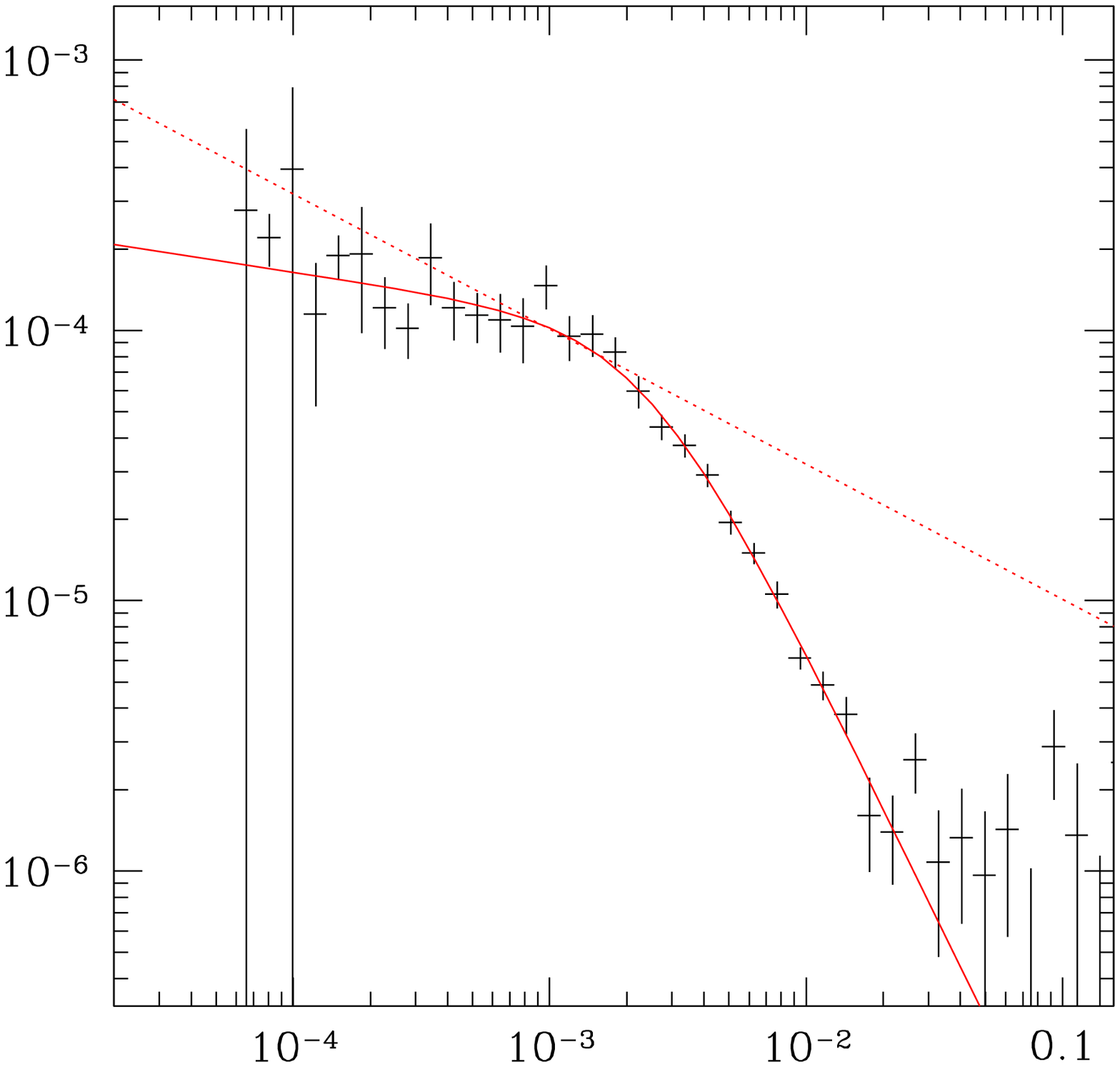}{Frequency, Hz}{Power$\times$Frequency,
    $(\sigma_x/\langle x \rangle)^2$}
  \caption{Power specrtum of optical flux variability of SS\,433
    during the periods of no accretion disk eclipses.}
  \label{fig:powspec_noecl}
\end{figure}

\begin{figure}
  \centering
  \smfigure{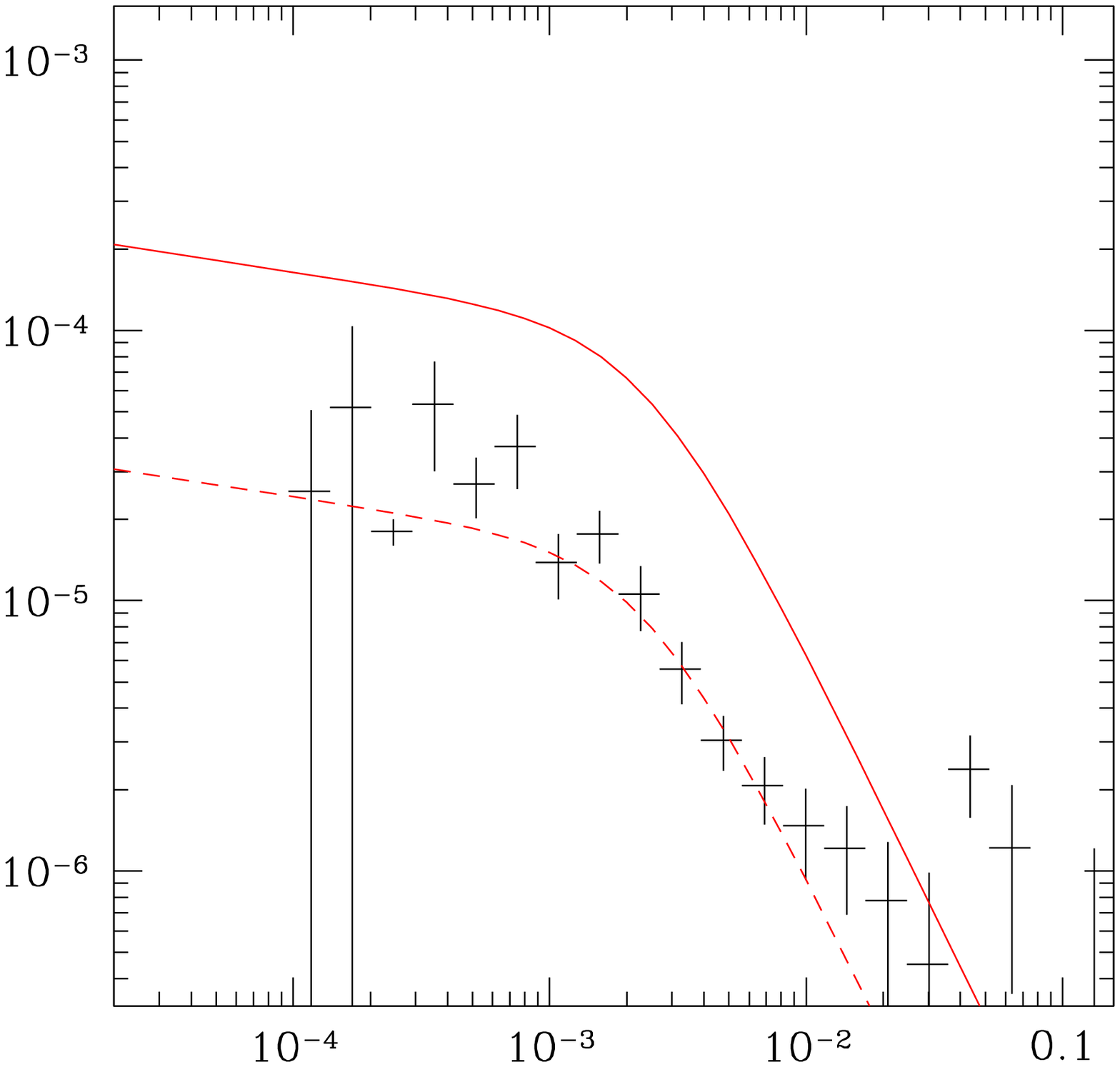}{Frequency, Hz}{Power$\times$Frequency,
    $(\sigma_x/\langle x \rangle)^2$}
  \caption{Power specrtum of optical flux variability of SS\,433
    during the eclipses of the accretion disk.}
  \label{fig:powspec_ecl}
\end{figure}

The power spectra of the optical variability of SS\,433 obtained from
our light-curve measurements are shown in Figs~\ref{fig:powspec_noecl}
and \ref{fig:powspec_ecl}. They were measured by averaging the
Lomb--Scargle periodograms \citep{lomb76,scargle82} that were
calculated from all continuous data segments and renormalized so that
to obtain the spectral power density of the variability in units of
the fractional RMS squared. In Figs.~\ref{fig:powspec_noecl} and
\ref{fig:powspec_ecl}, the spectral power density was additionally
multiplied by the frequency in order to compare the power in the
characteristic frequency intervals.

The constant corresponding to the Gaussian noise of the measurement
errors is subtracted from these power spectra. This constant was
determined from the power at frequencies $f>0.2$~Hz. Therefore, it was
assumed that the entire variability at these frequencies is explained
by the measurement errors. This should be close to reality with a good
accuracy, because even at frequencies $f>0.03$~Hz the power is
observed to be almost exactly constant and its value, from $3\times
10^{-4}$ to $4\times 10^{-4}$ for different light-curve segments,
agrees well with the constant power that was obtained during our
observations of nonvariable stars with close magnitudes (see
Fig.~\ref{fig:powspec_stars}).

Figure~\ref{fig:powspec_noecl} shows the power spectrum of the
variability of that part of the light curve, where the accretion-disk
eclipses were excluded, i.e., the light curve at the orbital phase
$|\varphi_{orb}|>0.1$. The best fit model $P(f) = f^{\alpha_1} [1 +
(f/f_0)^{2\alpha_2}]^{1/2}$, where $P(f)$ is the spectral power
density, is also shown in the Figure. The best fit parameters for this
model are $\alpha_1=-1.15\pm0.06$, $\alpha_2=-1.80\pm0.13$ and
$f_0=2.43\pm0.29\cdot10^{-3}$~Hz. The dotted line in the Figure shows
the power law with a slope of $-1.5$ that was obtained from the
observations of SS\,433 variability in the frequency range
$10^{-7}$--$10^{-2}$~Hz \citep{mikej06}.

The power-law slope of the power spectrum at high frequencies is
$\alpha_1 +\alpha_2 = -2.95$, i.e., it is steeper as compared to the
slope of the spectrum at low frequencies measured from our data
($\alpha_1=-1.15$) and to the slope $\alpha_1=-1.5$, which was
measured from the observations at lower frequencies by \cite{mikej06}.
Thus, our data reveal a break in the power spectrum at a frequency of
near $2.4\cdot10^{-3}$~Hz.

\begin{figure}
  \centering
  \smfigure{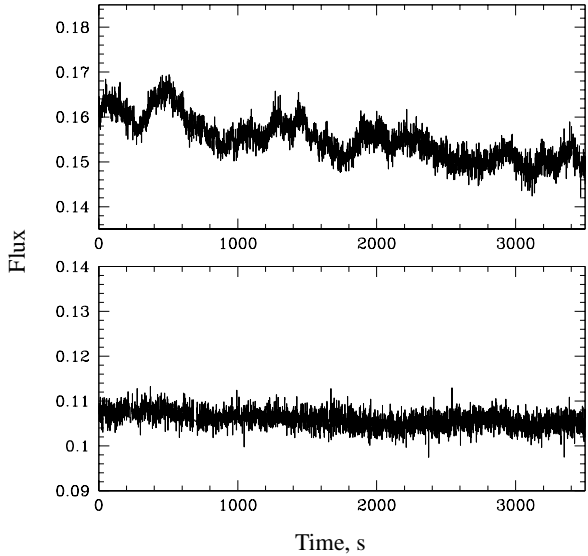}{Time, s}{Flux}
  \caption{The examples of SS\,433 light curves. Upper panel --- out
    of the eclipses of accretion disk, lower panel --- during the
    eclipse.}
  \label{fig:lc_noecl_ecl}
\end{figure}

In Fig.~\ref{fig:powspec_ecl} the power spectrum from our observations
during accretion-disk eclipses at an orbital phase
$|\varphi_{orb}|<0.1$ is shown. We see that the flux variability is
greatly reduced during disk eclipses. In this case, as far as can be
judged from the data, the shape of the power spectrum remains
approximately the same as that outside eclipses, while the
normalization decreases approximately by a factor of $4.5$. As an
example, in Fig.~\ref{fig:lc_noecl_ecl} we show the segments of the
light curves for SS 433 outside and during accretion-disk
eclipses. The drop in the optical variability amplitude during the
eclipse is seen in this Figure with the naked eye.

\begin{figure}
  \centering
  \smfigure{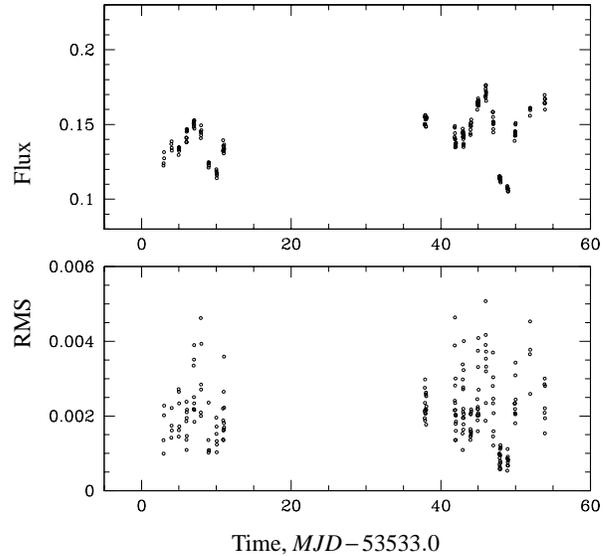}{Time, $MJD-53533.0$}
  {~~~~~~~~RMS ~~~~~~~~~~~~~~~~~~Flux~~~~~~}
  \caption{SS\,433 optical flux (upper panel) and its RMS at 1000~s
    time scale (lower panel).}
  \label{fig:lc_std}
\end{figure}

\begin{figure}
  \centering
  \smfigure{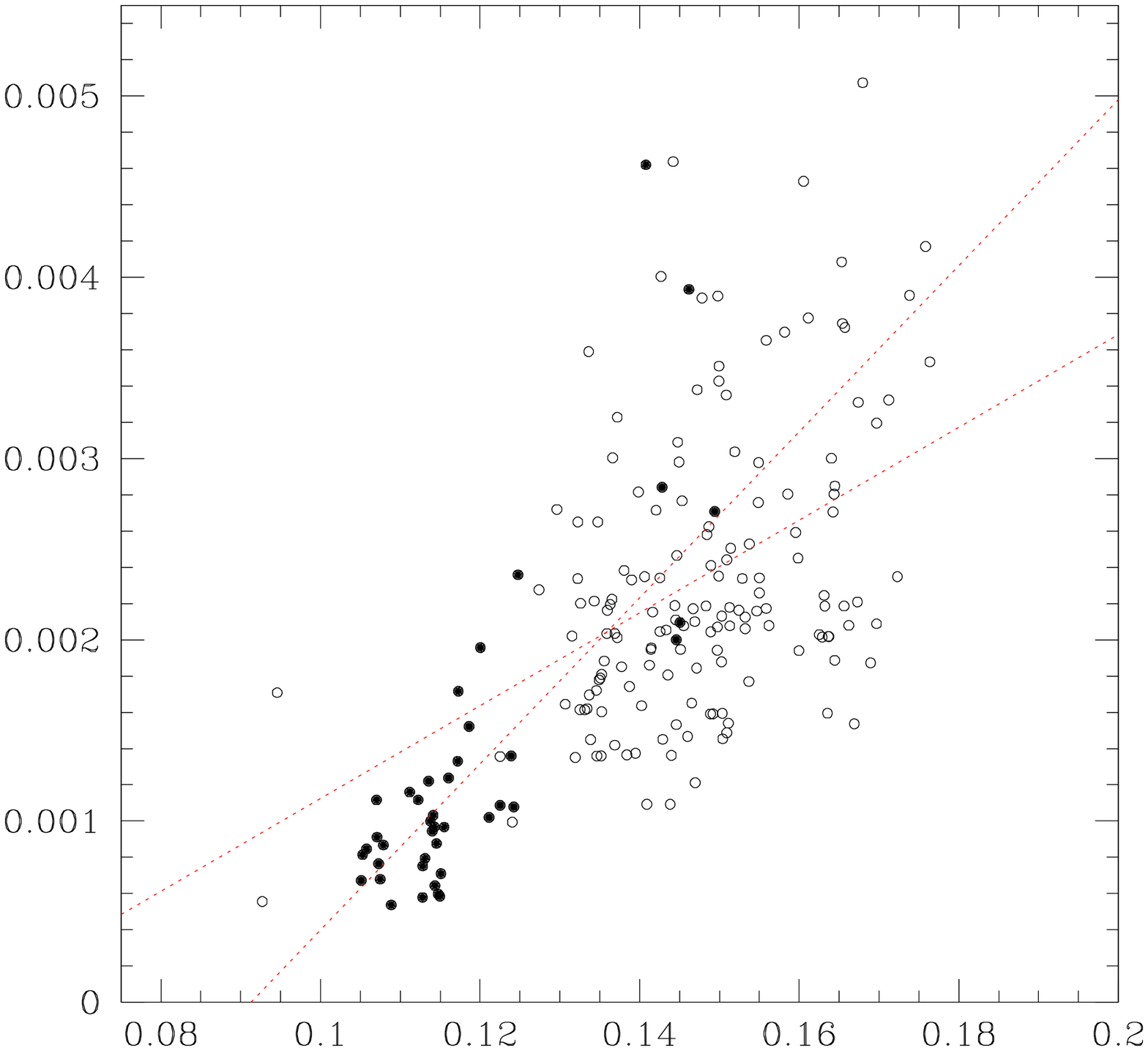}{Flux}
  {RMS}
  \caption{RMS -- optical flux relation for SS\,433 at 1000~s time
    scale. The filled circles indicate the measurements during eclipse
    at $\varphi_{orb}<0.1$}
  \label{fig:flux_std}
\end{figure}

In Figs.~\ref{fig:lc_std} and \ref{fig:flux_std} we show the relation
between the mean optical flux from SS 433 and its RMS at 1000~s time
scale. The filled circles in Fig.~\ref{fig:flux_std} indicate the
measurements when the orbital phase is $|\varphi_{orb}|<0.1$
(eclipse), while the open circles correspond to $|\varphi_{orb}|>0.1$
(out of eclipse). The measurements with fluxes about 0.14 at the
eclipse phase and with fluxes about 0.09 at the phase outside eclipse
correspond to the observations on July 19 and September 28, 2005,
respectively, when the orbital phase is close to a boundary value of
$\varphi_{orb}\approx0.9$. Thus, it follows from our data that the
eclipse ingress occurs at slightly different times at different
precession phases.

With the exception of these points, all of the remaining measurements
show not only a drop in optical flux during eclipse but also a
significant drop in the variability amplitude. In addition, we see
from Figs.~\ref{fig:lc_std} and \ref{fig:flux_std} that the optical
variability amplitude decreases not only during the eclipses but also
when the optical flux outside eclipse decreases. However, it turns out
that this occurs slightly more slowly than during eclipse. In
Fig.~\ref{fig:flux_std} the linear regressions for the data at fluxes
greater and less than 0.14 are shown. We see that these lines have
different slopes and the calculations of their slope estimates shows
that they are different at a confidence level of about $3\sigma$.

The change in the slope of the RMS -- mean flux relation in these two
cases is not something unexpected, because the accretion-disk eclipse
by the companion star changes the ratio of the fractions of the fluxes
from the variable and constant components in the total optical
emission recorded on Earth. Indeed, a direct proportionality between
the amplitude of aperiodic flux variations and the mean flux is
commonly observed in accreting binary systems \citep[see,
e.g.,][]{lyutyj87,uttley01}. If we could observe only one variable
emission component of the binary system (as it is observed, for
example, in the X-ray energy band), then one would expect the same
direct proportionalyty between RMS and mean flux to be observed at all
fluxes. However, the presence of a significant contribution of the
nonvariable flux from the companion star in the optical band changes
this relation. During the eclipse, the fractions of the variable and
constant fluxes undergo additional changes, causing the slope of the
RMS -- mean flux relation to change relative to the out-of-eclipse
case. We see from Fig.~\ref{fig:flux_std} that in the SS\,433 optical
emission during eclipses there is a nonvariable component with a flux
of about 0.09, which corresponds to a magnitude $m_R\approx13.2$.

\section{Variability smoothing and geometry of the emitting region}

As it was shown above, in the power spectrum of the optical
variability of SS\,433 obtained from our measurements out of the
accretion-disk eclipses there is a break at a frequency of about
$2.4\times 10^{-3}$~Hz. The presence of a break in the variability
power spectrum at high frequencies is expected, because the
variability must disappear at small time scales comparable to the time
delay of the photon arrival from different parts of the source of
variable emission. A frequency $f\approx2.4\cdot10^{-3}$~Hz
corresponds to a characteristic time $T=1/2\pi f \approx70$~s and to a
distance $R\approx2\cdot10^{12}$~cm.

The shape and exact position of the break in the power spectrum
depends on the geometry of the source of variable emission. For
example, if the entire variable emission originates on the surface of
a homogeneous sphere of radius $R$ that, in turn, is illuminated by a
central source, then the flux from an infinitely short flare of this
source will be recorded by a remote observer as a flare with a time
profile $\propto (1-t/Rc)$. If the central source is intrinsically
variable, then the flux recorded by a remote observer will correspond
to the flux from the central source convolved with this response
function, while the variability power spectrum of the flux recorded by
the observer will be the product of the power spectrum of the
intrinsic variability of central source and the square of the Fourier
transform of the above response function.

To get an idea of how the variability power spectrum for such a sphere
will appear, we performed the following simple simulation. At a
$10^4$~s time interval (which roughly corresponds to the time of the
continuous segment of observations in our data), we simulated a light
curve with a power-law variability power spectrum $P\propto f^{-1.5}$,
which was then convolved with the response function of the sphere
discussed above. After that, the power spectrum of this time serie was
obtained through the same procedure that was used to calculate the
power spectra when reducing the observational data.

\begin{figure}
  \centering
  \smfigure{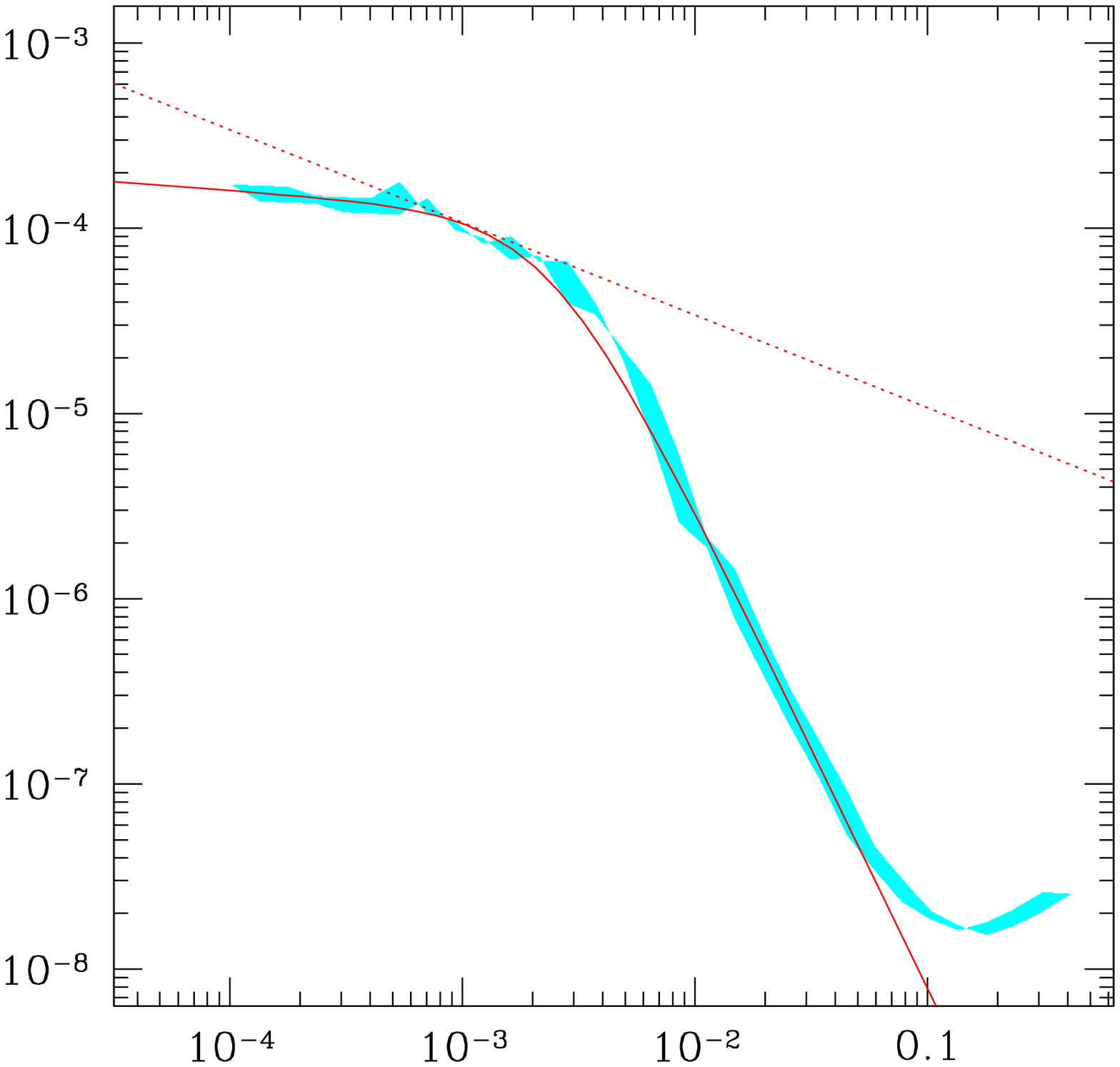}{Frequency, Hz}{Power$\times$Frequency,
    $(\sigma_x/\langle x \rangle)^2$}
  \caption{The filled strip shows the power spectrum of the
    variability of the observed optical emission from a homogeneous
    sphere with a size $R=2.46\cdot 10^{12}$~cm ($82.1$~s) whose
    surface emits variable radiation with a power spectrum $P\propto
    f^{-1.5}$ (dotted line). The solid curve indicates the model fit
    of this power spectrum.}
  \label{fig:powspec_mo}
\end{figure}

The power spectrum obtained in this way is shown in
Fig.~\ref{fig:powspec_mo} with the filled strip. The solid curve
indicates an analytical model fit of this power spectrum in the form
$P(f) = f^{\alpha_1} [1 + (f/f_0)^{2\alpha_2}]^{1/2}$ with parameters
$\alpha_1=-1.09$, $\alpha_2=-2.49$, $f_{0}=2.43\cdot 10^{-3}$~Hz. To
construct this power spectrum, we took the sphere size to be
$R=2.46\cdot 10^{12}$~cm (82.1~s). This size was chosen so that the
break occurred at a frequency $f\approx2.43\cdot10^{-3}$~Hz, as it is
observed in our data on the variability of SS\,433. The slope of this
power spectrum at high frequencies is found to be steeper than that
observed for SS\,433 (Fig.~\ref{fig:powspec_noecl}). This probably
suggests that the emitting region actually has a more complex
geometry.

In addition to the break at the characteristic frequency, the model
power spectrum in Fig.~\ref{fig:powspec_mo} also exhibits other
distortions related to the power spectrum measurement procedure ---
due to the red noise leakage and other effects, the power-law slope
decreases at low frequencies and an additional small noise component
is added at high frequencies, up to the Kotelnikov--Nyquist
frequency. The slope of the spectrum at frequencies near the lowest
measured frequency decreases, because the observed power spectrum is a
convolution of the true spectrum with the power spectrum of the window
function and, hence, the power at the lowest observed frequencies is
smeared toward higher frequencies.

Thus, the slope of the power spectrum at low frequencies is
underestimated in our measurements. For the slope of the power
spectrum at low frequencies to be properly measured, the corresponding
simulations must be performed. We are going to do this in
future. Nevertheless, these distortions should not affect strongly the
measurement of the break frequency in the power spectrum.

From the above example we see that, given a specific model of the
variable optical emission source in SS\,433, from the observed power
spectrum we can determine its geometrical size with a good accuracy
and can even estimate the level of applicability of a particular
geometrical model. For the most plausible models of the source, we are
going to do this in future as well.

\section{Discussion}

During our optical observations of SS\,433, we used the CCD setup when
not the image but only one-dimensional data line was read out after
each exposure. This allowed to improve the time resolution of the
photometric measurements to $\sim 1$~s. At the same time, the
signal-to-noise ratio of optical flux measurements remains high ---
the scatter of the optical flux measurements for SS\,433, whose
magnitude is $m_R\approx12.5$, constitutes only about two percent of
the flux, i.e., it turns out to be comparable to the expected scatter
due to Poissonian photon noise.

It is important that in these observations the possibility of
simultaneous measurements of the fluxes from at least two stars (the
object and the reference star) and obtaining diffrential flux
measurements for the object is retained. Through the observations of
nonvariable stars, we showed that this allows to remove almost
completely the influence of atmospheric turbulence on the photometric
measurements. Therefore, the data obtained by this technique are well
suited for studying the \emph{aperiodic} variability of various
objects. We applied this observing technique recently in our study of
the inner regions of the accretion disks in a sample of intermediate
polars through the observations of their optical variability
\citep{mikej10}.

In this paper, we obtained and studied a large volume of data on the
optical variability of SS\,433. The observations were performed during
more than twenty nights, mainly in the summer of 2005. For our
observations, we chose the precession phase when the accretion disk
was maximally turned toward the observer, because the variability of
the source at this phase is at maximum as well.

The power spectrum of the optical variability in SS\,433 obtained from
our light curve measurements out of accretion disk eclipses exhibits a
break at a frequency of about $2.4\cdot10^{-3}$~Hz. At higher
frequencies the variability power falls down rapidly; the power-law
slope of the power spectrum is $\approx-2.95$. In spite of this, a
statistically significant optical variability of SS\,433 is detected
up to frequencies $\sim3\cdot10^{-2}$~Hz, with its relative amplitude
at these frequencies being only of order $\sim 0.1$\%.

We propose to explain the presence of a break in the variability power
spectrum of SS\,433 as a manifestation of the smoothing of the
intrinsic variability of the source due to its finite size. Under this
assumption our measurement of the break frequency gives a new,
independent of other methods, measurement of the size of the variable
optical emission source in SS\,433. A frequency
$f\approx2.4\cdot10^{-3}$~Hz corresponds to a characteristic
time$T=1/2\pi f \approx70$~s and a size $R\approx2\cdot10^{12}$~cm. As
we show above, a more accurate measurement of the size depends on the
geometrical model of the source. However, given a specific model, we
can determine its size from our data with good accuracy and can
estimate the degree of the applicability of this partucilar
geometrical model.

Our measurement of the size of the variable optical emission source
turns out to be close to the estimate of the size of the hot optical
emission source, $R\approx2\cdot10^{12}$~cm, obtained by modeling the
energy distribution based on broadband photometry in the optical and
ultraviolet wavelength bands \citep{dolan97}. A slightly smaller size,
$R\approx(0.06\div0.09)a \approx 3\cdot10^{11}$~cm is obtained from
the measurement of the distance at which an X-ray emitting
relativistic jet emerges from an opaque cloud of matter surrounding
the source \citep{filippova06}. This may suggest that the optical hot
variable source is asymmetric in projection onto the plane of the sky.

Our observations also show that the amplitude of the variability in
SS\,433 decreases sharply during accretion disk eclipses, but it does
not disappear completely. This suggests that the size of the variable
optical emission source is comparable to that of the normal star,
which, consequently, should also be $R_O\approx 2\cdot10^{12}$~cm
$\approx 30 R_\odot$. The eclipse occurs as if, apart from the
variable source, there is also a nonvariable source with a magnitude
$m_R\approx13.2$. This magnitude corrected for extinction $A_V \approx
8.4$ \citep{dolan97} corresponds to the absolute magnitude $M_R\approx
-7.2$. Note that about a half of the $R$-band emission originates in
the wind around the binary system \citep{dolan97,gech98a}.

Our estimates of the radius of the normal star and the absolute
magnitude of nonvariable component in the source are consistent with
the assumption that the normal star is an A-type supergiant, as it was
suggested previously from photometric measurements during eclipses
\citep{cherepaschuk82} and from the observations of spectral lines of
the normal star \citep{hillwig04,hillwig08}. The radius of the normal
star also agrees well with the estimate obtained from the observations
of eclipses in X-rays \citep{filippova06}.

On the whole, the geometrical sizes of the source determined by
studying the optical variability of SS\,433 agree well with those
measured previously by other methods. It should be emphasized that our
measurements were obtained with a new, completely independent
technique that uses a new physical phenomenon for such studies --- the
smoothing of the flux variability due to a finite light travel time
across the size of the emitting system. Our results suggest that the
existing views of the sizes of the system are mostly correct and
contain no large systematic errors. In addition, we can conclude that
the measurements of the aperiodic variability of X-ray binaries in the
optical band is an efficient method of studying the geometry of the
emitting regions in such systems.

\acknowledgements

We are grateful to TUBITAK National Observatory (TUG, Turkey), the
Space Research Institute of the Russian Academy of Sciences (IKI RAN),
and the Kazan State University for support in using the
Russian-Turkish 1.5-m telescope (RTT-150). This work was supported by
the Russian Foundation for Basic Research (project nos. 07-02-01004,
08-02-00974, 09-02-12384-ofi\_m, 10-02-01442, 10-02-01145, 10-02-00492,
10-02-91223-ST\_a), the Program for Support of Leading Scientific
Schools of the Russian Federation (Nsh-5069.2010.2), and the Programs
of the Russian Academy of Sciences P-19 and OPhN-16.

\end{document}